\documentclass[12pt]{article}
\usepackage{makeidx}
\usepackage{multirow}
\usepackage{multicol}
\usepackage[dvipsnames,svgnames]{xcolor}
\usepackage{graphicx}
\usepackage{epstopdf}
\usepackage{ulem}
\usepackage{enumerate}
\usepackage{relsize}
\usepackage{amsmath}
\usepackage{amssymb}
\usepackage[bookmarks]{hyperref}
\hypersetup{
    bookmarks=true,         % show bookmarks bar?
    unicode=false,          % non-Latin characters in Acrobat's bookmarks
    pdftoolbar=true,        % show Acrobat's toolbar?
    pdfmenubar=true,        % show Acrobat's menu?
    pdffitwindow=true,      % page fit to window when opened
    pdftitle={My title},    % title
    pdfauthor={Author},     % author
    pdfsubject={Subject},   % subject of the document
    pdfnewwindow=true,      % links in new window
    pdfkeywords={keywords}, % list of keywords
    colorlinks=true,        % false: boxed links; true: colored links
    linkcolor=blue,         % color of internal links
    citecolor=blue,        % color of links to bibliography
    filecolor=magenta,      % color of file links
    urlcolor=cyan           % color of external links
}
\usepackage{geometry}
\usepackage{framed}    % to make framed boxes
\usepackage{scalefnt}
\usepackage{alltt}
\usepackage{setspace}

\def\bsb{\boldsymbol \beta}

\def\bsm{\boldsymbol \mu}
\def\bsS{\boldsymbol \Sigma}

\usepackage{natbib}
\begin{document}
\thispagestyle{empty}

\title{Copula Modeling of Multivariate Longitudinal Data with Dropout}

\author{Edward W. Frees, Catalina Bolanc\'{e}, Montserrat Guillen, Emiliano A. Valdez
\thanks{
Edward W. Frees, University of Wisconsin-Madison.
Catalina Bolanc\'{e}, Universitat de Barcelona.
Montserrat Guillen, Universitat de Barcelona.
Emiliano A. Valdez, University of Connecticut.
}}

\date{\today}

\maketitle

\subsubsection*{Summary}

Joint multivariate longitudinal and time-to-event data are gaining increasing attention in the biomedical sciences where subjects are followed over time to monitor the progress of a disease or medical condition. In the insurance context, claims outcomes may be related to a policyholder's dropout or decision to lapse a policy. This paper introduces a generalized method of moments technique to estimate dependence parameters where associations are represented using copulas. A simulation study demonstrates the viability of the approach. The paper describes how the joint model provides new information that insurers can use to better manage their portfolios of risks using illustrative data from a Spanish insurer.

\bigskip

\noindent\textit{Keywords and Phrases:} Dependence modeling, generalized method of moments, joint longitudinal and time-to-event, insurance lapsation, repeated measurements, Tweedie distribution

\newpage

\scalefont{0.6}
%\tableofcontents
\scalefont{1.66667}

\begin{spacing}{1.8}

\newpage

\section{Introduction}\label{S:Intro}

\textbf{Multivariate longitudinal modeling.} Modeling of repeated observations of a subject over time, the topic of longitudinal or panel data, has long been an important tool in the biomedical and social sciences. Naturally, the field initially focused attention on the analysis of data in which a single outcome is analyzed. Subsequently, the desire to handle multivariate outcomes has gained prominence, as emphasized in a review by \cite{verbeke2014analysis}. For example, one might be interested in studying hearing ability changes during aging based on longitudinal measurements of subjects. Multivariate outcomes arise naturally through multiple responses to hearing tests at various frequencies for both the left and right ear. As another example, in this paper we study multivariate insurance losses. Specifically, we consider a sample of individuals, followed over time, who have purchased more than one type of insurance policy with claims arising from each contract. To illustrate, we consider people who purchased policies that cover insured losses on an automobile and a home.

%The interest is in the joint behavior of these two types of claims.

As emphasized by \cite{verbeke2014analysis}, a multivariate modeling strategy is more complex than analyzing univariate outcomes but also provides important insights not available with univariate analysis. For example, with a multivariate model, a researcher may wish to assess the relation between some covariate and all outcomes simultaneously. Further, multivariate modeling is needed to understand associations among outcomes and how these relationships evolve over time.

%Verbeke et al. (2014) categorize multivariate longitudinal models into four types. One broad type is based on observable outcomes, subdivided into marginal models and models built on conditioning. The other three types are different types of latent variable models where associations among outcomes are build on unobserved, latent, variables that are common to outcomes. As described in the following, the approach of this paper is to focus on the marginal observable outcomes approach.

\smallskip

\textbf{Joint models of longitudinal and time-to-event data.} A well known limitation of longitudinal data models, with either univariate or multivariate outcomes, is the bias caused by event times that dictate outcome availability where the event is related to the outcome being studied.  As an example from the biomedical field, outcomes can be a function of patient medical expenditures which are associated with survival. Intuitively, it seems plausible that sicker patients are more likely to die (and thus exit a study) and to incur higher medical costs, an outcome of interest, c.f., \cite{liu2009joint}.  To address this, joint models of longitudinal and time-to-event data have been developed in the biomedical literature; for example, \cite{tsiatis2004joint} and \cite{ibrahim2010basic} review the foundations. A more recent overview by \cite{papageorgiou2019overview} describes the explosion of the field including the development of several statistical packages within {\tt{R}} (R Core Team 2018), SAS/STAT, Stata, WinBUGS, and JAGS  that enable innovative applications of the methodology.

\smallskip

\textbf{Joint multivariate longitudinal and time-to-event data.} Many applications of joint models for longitudinal and time-to-event data appear in settings where subjects are followed over time to monitor the progress of a disease or medical condition. That progression is typically evaluated via repeated measurements of biomarkers pertinent to the disease, subject to an event of interest such as death or intervention, and there may be many such biomarkers. As emphasized by \cite{hickey2016joint}, the value of multivariate modeling is particularly well established for applications involving personalized medicine, where methodological tools are tailored to the individual characteristics of their patients.

\smallskip

\textbf{Modeling associations.} \cite{elashoff2017joint} describe two basic approaches for modeling the association between longitudinal data and time-to-events, both involving conditioning. The first is the selection model, where the joint distribution between outcomes of interest and an event is factored into a distribution of outcomes and the distribution of events given the outcomes. The second basic approach is the pattern mixture model, where the factorization is reversed. That is, the joint distribution is factored into the distribution of events and the distribution of outcomes given the events.

Other sources, such as \cite{njagi2014characterization}, think of the ``shared parameter model'' as a separate category (\cite{elashoff2017joint} include this approach as a selection model). In the shared parameter model, one or more latent variables may induce an association between survival/drop-out and outcomes (e.g., costs). A strength of this approach is that latent variables are intuitively appealing, one can ascribe attributes to them (e.g., ``frailty'') to help interpret dependence.

%A limitation of this approach is that the marginal model structure is not maintained (in the non-linear case), making it more difficult to use data for model identification. As an alternative, we explore copula modeling.

\smallskip

\textbf{Insurance data and copula modeling}. This paper extends the joint multivariate longitudinal and time-to-event data modeling literature in two ways. First, it considers insurance, a new field of application. This new application suggests concerns that are not commonly seen in biomedical applications, including marginal outcome distributions that are hybrid combinations of continuous and discrete components and larger sample sizes (suggesting the need for alternative computational approaches). Considerations of this new application area lead naturally to the introduction of copula models as a tool to understand associations. As noted in the review by \cite{verbeke2014analysis}, copula models have been utilized in the multivariate longitudinal data literature although it is not widely used in biomedical applications; early contributions include
\cite{meester1994parametric},
\cite{lambert1996modelling},
\cite{lambert2002copula}, and
\cite{frees2005credibility}. Second, the copula approach for modeling dependencies has not been explored for joint models of longitudinal and time-to-event data.

As we argue in this paper, there are several strengths of copula models compared to alternatives. First, copulas allow for a unified model of associations while preserving marginal distributions. In complex applications like insurance where marginal distributions are hybrid combinations of continuous and discrete components, this allows analysts to use traditional diagnostic techniques to specify marginal distributions. Second, there is no need to make assumptions about conditioning when specifying joint distributions. In the insurance context, we will think about outcomes of interest as insurance claims and events as lapsation, or dropout, from insurance policies. In advance, we do not know when an insurance claim causes a policyholder to lapse a policy, or whether the decision to lapse influences the outcome of a claims. With a copula approach, we model the observed dependency directly without the need for making decisions about conditioning. Third, we argue that the computational complexity of the copula modeling, with the new estimation procedures introduced in this paper, allow us to handle some large problems that would difficult, if not infeasible, using alternative available strategies.

\smallskip

%\subsection*{Dependence and Foundations of Insurance}\label{S:InsuranceDependence}

\label{S:CopulaModeling}
\textbf{Copula regression modeling.} For notation, consider the joint distribution of $p$ outcomes, $Y_1, \ldots, Y_p$. Copulas provide a general tool for modeling dependencies and so express the joint distribution as
\begin{equation}\label{E:MainCop}
F(y_1, \ldots, y_p) = \Pr(Y_1 \le y_1, \ldots, Y_p \le y_p) = C\left(F_{1}(y_1), \ldots, F_p(y_p) \right).
\end{equation}
Here, $F_1(y_1) = \Pr (Y_1 \le y_1)$ is known as the marginal distribution of $Y_1$ and similarly for the other variables. We assume that there is an associated set of explanatory variables, $\mathbf{x}$, that is available to calibrate the marginal distributions. When the distribution of the marginal distributions depends on covariates, we refer to this as a \textit{copula regression model}. If we think of $Y_1, \ldots, Y_m$ as realizations from the same subject over time (we use $m$ for the number of time points), then this framework reduces to longitudinal data. In the same sense, if a realization at a time point consists of multiple outcomes, then this is multivariate longitudinal data.

Copula regression modeling is ideally suited for applications where there are many variables available to explain outcomes (the regression portion) and where structural dependence among outcomes is critical (the copula portion). Compared to other multivariate techniques, copulas are particularly suitable in insurance applications because there is a lack of theory to support specification of a dependence structure and data-driven methods, such as copula modeling, fare well.

\smallskip

\textbf{Estimation using generalized method of moments.} Copula regression modeling is introduced in greater detail in Section \ref{S:CopRegression} with motivating insurance applications in \textit{Online Supplement} 1. Traditionally, a barrier to implementing these models in high-dimensional cases (large $p$) has been the presence of discreteness which substantially increases the computational burden. In this paper, we show how to use the generalized method of moments (GMM), a type of estimating equations approach, to estimate association parameters of this model. Unlike traditional treatments, we emphasize examples where outcome variables may be a hybrid combination of continuous and discrete components. As will be described, this approach provides an alternative to vine copula models that have been developed recently.

\smallskip

\textbf{Multivariate longitudinal models with dropout}. Starting with $Y_1, \ldots, Y_m$ vectors, we now  allow $Y_t$ to represent a multivariate outcome and an indicator for dropout in Section \ref{S:LapseModel}. This section is an extension, and not a special case, of Section \ref{S:CopRegression} because of potential dependencies between the outcomes and the observation process. To provide intuition, Section \ref{S:LapseIndependence} focuses on the special case of temporal independence (but still accounting for dependency between dropout and outcomes during the same time period).

Section \ref{S:Empirical} describes an empirical application where we examine outcomes and dropout (lapsation), tracking 40,284 clients over five years. Section \ref{S:Conclusion} concludes.

\section{Copula Regression Modeling and GMM Estimation}\label{S:CopRegression}

In copula regression applications, it is common to use the \textit{inference for margins} (IFM) procedure, a two stage estimation algorithm due to \cite{joe2005asymptotic}. The first stage maximizes the likelihood from marginals, and the second stage maximizes the likelihood of dependence parameters with parameters in the marginals held fixed from the first stage. For insurance applications of interest to us, this general model fitting strategy works well and will be utilized in the subsequent development.

\subsection{Hybrid Distributions}\label{S:PairwiseDist}

Assume that the random variables may have both discrete and continuous components. The need for handling both discrete and continuous components was emphasized by \cite{song2009joint} in the copula regression literature. They referred to this combination as a ``mixture'' of discrete and continuous components. In insurance and many other fields, the term ``mixture'' is used for distributions with different sub-populations that are combined using latent variables. So, we prefer to refer to this as a ``hybrid'' combination of discrete and continuous components to avoid confusion with mixture distributions. For a random variable $Y$, let $y^d$ represent a mass point ($d$ for discrete) and let $y^c$ represent a point of continuity (where the density is positive).

\subsubsection*{Likelihood}

We now give a general expression for the likelihood. To do so, assume that the first $d$ arguments $y_1, \ldots, y_d$ represent mass points. Further assume that the other arguments $y_{d+1}, \ldots, y_p$ represent points of continuity. Then, the likelihood corresponding to the distribution function in equation \eqref{E:MainCop} can be expressed as

\begin{eqnarray}\label{E:MainLikelihood}
&&f(y_1, \ldots, y_d, y_{d+1}, \ldots, y_p) = \nonumber \\
&&\ \ \ \ \sum_{i_1=0}^1 \cdots  \sum_{i_d=0}^1 \left(-1\right)^{i_1 + \cdots + i_d}
C_{d+1, \ldots, p} \left(F_{1}(y_1^{(i_1)}), \ldots, F_{d}(y_d^{(i_d)}), F_{d+1}(y_{d+1}), \ldots, F_{p}(y_{p})  \right) \nonumber\\
&&\ \ \ \  \ \ \ \ \ \times \prod_{j=d+1}^p f_{j}(y_{j}) ,
\end{eqnarray}
where $y^{(i)} = y$ if $i=0$ and $y^{(i)} = y-$ if $i=1$. The notation $y-$ means evaluate $y$ as a left-hand limit. See, for example, \cite{song2009joint}, equation (9). Here, $C_{d+1, \ldots, p}$ is the partial derivative of the copula with respect to the arguments in the $d+1, \dots, p$ positions. In common elliptical families, evaluating partial derivatives of a copula is a tedious, yet straightforward, task. As we will see, a large number of variables with discrete components ($d$ in equation \eqref{E:MainLikelihood}) represents a source of difficulty when numerically evaluating likelihoods.

\subsubsection*{Pairwise Distributions}\label{S:PairwiseDist}
Particularly for discrete or hybrid outcomes, direct estimation using maximum likelihood (or IFM) can be difficult because the copula distribution function may require a high-dimensional evaluation of an integral for each observation. A generally available alternative is to examine the information from subsets of random variables. For example, focusing on pairs, consider the corresponding bivariate distribution function
$$
F_{jk}(y_j, y_k) = C\left( \infty, \ldots, \infty ,F_{j}(y_j), \ldots, F_{k}(y_k), \infty, \ldots, \infty \right)= C^{(jk)}\left(F_{j}(y_j),F_{k}(y_k)\right).
$$
Although this expression for the joint distribution function $F_{jk}$ is broadly applicable, it is particular useful for copulas in the elliptical family. If we specify $C$ to be an elliptical copula with association matrix $\bsS$, then $C^{(jk)}$ is from the same elliptical family with association parameter $\bsS_{jk}$, corresponding to the $j$th row and $k$th column of $\bsS$. This is the specification used in this paper. For notational purposes, we henceforth drop the superscripts on the bivariate copula and hope that the context makes the definition clear.

Suppose that we wish to evaluate the likelihood of $Y_j$ at mass point $y_j^d$ and of $Y_k$ at point of continuity $y_k^c$. Then, the joint distribution function has a hybrid probability density/mass function of the form:
$$\begin{array}{ll}
f_{jk}(y_j^d,y_k^c) &=
\partial_2 \Pr(Y_{j} = y_j^d, Y_{k} \le y_k^c)  \\
&=\partial_2 \left\{ \Pr(Y_j \le y_j^d, Y_k \le y_k^c) -\Pr(Y_j \le y_j^d-, Y_k \le y_k^c) \right\}\\
&=\partial_2 \left\{C \left( F_j(y_j^d),F_k(y_k^c)\right) -C \left( F_j(y_j^d-),F_k(y_k^c)\right) \right\}\\
&= \left\{C_2 \left( F_j(y_j^d),F_k(y_k^c)\right) -C_2 \left( F_j(y_j^d-),F_k(y_k^c)\right) \right\} f_k(y_k^c) .
\end{array}
$$
Here, $C_2$ represents the partial derivative of the copula $C$ with respect to the second argument. For the likelihood, we need to consider two other cases, where $y_j$ and $y_k$ are both points of continuity or both discrete points. These cases are more familiar to most readers; details are available in our \textit{Online Supplement} 3.6.

\subsubsection*{Example: Two Tweedie Variables}

In insurance, it is common to refer to a random variable with a mass at zero and a continuous density over the positive reals as a ``Tweedie'' random variable (corresponding to a Tweedie distribution). Suppose that both $Y_j$ and $Y_k$ are Tweedie random variables. The joint distribution function has a hybrid probability density/mass function of the form:
$$
f_{jk}(y_j,y_k) = \left\{
\begin{array}{ll}
\Pr(Y_{j} =0, Y_{k}=0) = F_{jk}(0,0) \ \ \ \       & y_j=0,y_k=0 \\
C_1 \left(F_j(y_j), F_k(0) \right) f_j(y_j)        & y_j>0,y_k=0 \\
C_2 \left( F_j(0),F_k(y_k)\right) f_k(y_k)         & y_j=0,y_k>0\\
c\left( F_j(y_j),F_k(y_k)\right)f_j(y_j) f_k(y_k)  & y_j>0,y_k>0 .
\end{array} \right.
$$
Here, $C_1$ represents the partial derivative of the copula $C$ with respect to the first argument and $c$ is the corresponding density. To illustrate, when both observations are 0, then the likelihood $F_{jk}(0,0) = C\left(F_{j}(0),F_{k}(0) \right)$ requires evaluation of the distribution function $C$. With a Gaussian copula, this is a two-dimensional integral. In the same way, with $m$ years of data, an observation that has no claims (all 0's) in all years requires a $m$-dimensional integration. For datasets that have tens of thousands of observations, this becomes cumbersome even when $m$ is small, e.g., $m=5$ as in our application. This is one motivation for utilizing pairwise distributions in this paper.

\subsection{Pairwise Likelihood-Based Inference}\label{S:PLikeProcedure}

When the number of discrete components is small, then full maximum likelihood estimation typically is feasible, c.f., \cite{song2009joint}. For copula likelihoods, the presence of a large number of discrete components means high-dimensional integration is needed to evaluate the full likelihood which can be computationally prohibitive. Instead, analysts look to (weighted sums of) low dimensional marginals that have almost as much information as a full likelihood and that are much easier to evaluate. Elliptical copulas naturally lend themselves to this approach as structure is preserved when examining smaller dimensions. As described in \cite{joe2014dependence}, Section 5.6, the strategy of basing inference on low-dimensional marginals is a special case of the ``composite likelihood'' approach.

As one special case, we have already mentioned the IFM technique. As another, consider a ``pairwise'' (logarithmic) likelihood that has the form
$$
L = \sum_{i=1}^n \left\{ \sum_{j=1}^{p-1} \sum_{k=j+1}^p  \ln f_{ijk}(y_{ij},y_{ik}) \right\} .
$$
This assumes independence among policyholders but permits dependence among risks. The pairwise likelihood procedure works well and has been used successfully in connection with copulas, c.f., \cite{shi2014insurance} and \cite{jiryaie2016gaussian}.

We can express pairwise likelihood procedure in terms of an estimating, or inference function, equation that will allow us to see why pairwise likelihood works. Specifically, assume that the parameters of the marginal distribution have been fit. Now, we wish to estimate a vector of association parameters $\theta$, a $r \times 1$ vector, that captures dependencies. To estimate $\theta$, we define the score function
\begin{equation}\label{E:Score}
g_{\theta,ijk}(Y_{ij},Y_{ik}) = \partial_{\theta}  \ln f_{ijk}(Y_{ij},Y_{ik})   .
\end{equation}
This is a mean zero random vector that contains information about $\theta$. It is an unbiased estimator (of zero) and can be used in an estimating equation. Under mild regularity conditions, c.f. \cite{song2007correlated}, minimizing the pairwise likelihood yields the same estimate as solving the estimating equation
$$
\sum_{i=1}^n \left\{ \sum_{j=1}^{p-1} \sum_{k=j+1}^p  g_{\theta,ijk}(Y_{ij},Y_{ik})) \right\} =0 .
$$
Denote this estimator as $\theta_{PL}$.

\subsection{Generalized Method of Moments Procedure}

As described in \cite{joe2014dependence}, Section 5.5, much of theory needed to justify composite likelihood methods in a copula setting can be provided through estimating equations. We utilize an estimating equation technique that is common in econometrics, \textit{generalized method of moments}, denoted by the acronym \textit{GMM}. Compared to classic estimating equation approaches, this technique has the advantage that a large number of equations may be used and that the resulting estimators enjoy certain optimality properties. The \textit{GMM} procedure is broadly used although applications in the copula context have been limited. For example, \cite{prokhorov2009likelihood} consider \textit{GMM} and copulas in a longitudinal setting although not with discrete outcomes.

\subsubsection{GMM Procedure}\label{S:GMMProcedure}

Using equation \eqref{E:Score}, we combine several scores from the $i$th risk as
\begin{equation}\label{E:ScoreVec}
g_{\theta,i}(Y_{i1}, \ldots, Y_{ip}) =
\left\{\begin{array}{c}
        g_{\theta,i12}(Y_{i1},Y_{i2}) \\
        \vdots \\
        g_{\theta,i1p}(Y_{i1},Y_{ip}) \\
        \vdots \\
        g_{\theta,i,p-1,p}(Y_{i,p-1},Y_{ip})
      \end{array} \right.
\end{equation}
a column vector with $r \left(^p_2\right)$ rows. The sum of these statistics for a sample of size $n$ is
$g_{\theta} = \sum_{i=1}^n g_{\theta,i}(Y_{i1}, \ldots, Y_{ip}).$ Although $g_{\theta}$ is a mean zero vector containing information about $\theta$, the number of elements in $g_{\theta}$ exceeds the number of parameters and so we use \textit{GMM} to estimate the parameters. Specifically, the \textit{GMM} estimator of $\theta$, say $\theta_{GMM}$, is the minimizer of the expression $g_{\theta} \left( \mathrm{Var~} g_{\hat{\theta}} \right)^{-1} g_{\theta}^{\prime}$. To implement this, we use the plug-in estimator of the variance
\begin{equation}\label{E:ScoreMatrix}
\mathrm{\widehat{Var}}~g_{\hat{\theta}} = \frac{1}{n} \sum_{i=1}^n g_{\hat{\theta},i}(Y_{i1}, \ldots, Y_{ip})~g_{\hat{\theta},i}(Y_{i1}, \ldots, Y_{ip})^{\prime}.
\end{equation}
This plug-in estimator requires $\hat{\theta}$, a consistent estimator of $\theta$. We use $\theta_{PL}$, defined in Section \ref{S:PLikeProcedure}.

For asymptotic variances, we use the gradient $G_{\theta}= \mathrm{E~} \frac{\partial}{\partial \theta^{\prime}} g_{\theta}$, a matrix of dimension $\left(r \left(^p_2\right)\right) \times r$. Then, following the usual asymptotic theory (see, for example, \cite{wooldridge2010econometric}, Chapter 14), we have that $\theta_{GMM}$ is asymptotically normal with mean $\theta$ and variance
$n^{-1} \left( G_{\hat{\theta}}^{\prime} \left( \mathrm{Var~} g_{\hat{\theta}} \right)^{-1} G_{\hat{\theta}} \right)^{-1} .$
Further,
$n^{-1} g_{\hat{\theta}}^{\prime} \left( \mathrm{Var~} g_{\hat{\theta}} \right)^{-1} g_{\hat{\theta}} $
has a limiting chi-square distribution with $r \left(^p_2\right)-r$ degrees of freedom.

\subsubsection{Evaluation of GMM Scores}\label{S:GMMScoreEval}

We now evaluate the scores. For discrete $y_j^d$ and continuous $y_k^c$ outcomes, we have
$$\begin{array}{cl}
g_{\theta,ijk}(y_j^d,y_k^c)
&= \partial_{\theta}  \ln \left[
 \left\{C_2 \left( F_j(y_j^d),F_k(y_k^c)\right) -C_2 \left( F_j(y_j^d-),F_k(y_k^c)\right) \right\} f_k(y_k^c) \right] \\
&= \mathlarger{ \frac{\partial_{\theta} \left\{ C_2 \left( F_j(y_j^d),F_k(y_k^c)\right) -C_2 \left( F_j(y_j^d-),F_k(y_k^c)\right) \right\}}
{C_2 \left( F_j(y_j^d),F_k(y_k^c)\right) -C_2 \left( F_j(y_j^d-),F_k(y_k^c)\right)} }
\end{array}$$

\end{spacing}

\noindent For two discrete outcomes, the score can be expressed as
\scalefont{1.11111}
$$\begin{array}{cl}
g_{\theta,ijk}(y_j^d,y_k^d) &= \partial_{\theta}  \ln f_{ijk}(y_j^d,y_k^d) \\
& = { \frac{\partial_{\theta} \left\{C \left( F_j(y_j^d),F_k(y_k^d)\right) -C \left( F_j(y_j^d-),F_k(y_k^d)\right) -C \left( F_j(y_j^d),F_k(y_k^d-)\right) +C \left( F_j(y_j^d-),F_k(y_k^d-)\right)\right\}}
{C \left( F_j(y_j^d),F_k(y_k^d)\right) -C \left( F_j(y_j^d-),F_k(y_k^d)\right) -C \left( F_j(y_j^d),F_k(y_k^d-)\right) +C \left( F_j(y_j^d-),F_k(y_k^d-)\right)} }.
\end{array}$$
\scalefont{0.9}
\begin{spacing}{1.8}

\noindent For two continuous outcomes, $y_j^c$ and $y_k^c$, we have
$$\begin{array}{cl}
g_{\theta,ijk}(y_j^c,y_k^c)&= \partial_{\theta}  \ln \left[  c(F_{Y_{ij}}(y_j^c), F_{Y_{ik}}(y_k^c)) f_{ij}(y_j^c) f_{ik}(y_k^c) \right]  \\% ~ \\
&= \mathlarger{ \frac{\partial_{\theta}  ~  c(F_{Y_{ij}}(y_j^c), F_{Y_{ik}}(y_k^c))}
                         { c(F_{Y_{ij}}(y_j^c), F_{Y_{ik}}(y_k^c))} }.
\end{array}$$

In the Appendix and \textit{Online Supplement} 2, we give explicit formulas for Gaussian copula derivatives. An advantage of restricting ourselves to pairwise distributions is that most of the functions are available from \cite{schepsmeier2012web, schepsmeier2014derivatives}. We use an additional relationship, from \cite{plackett1954reduction}, that is developed in the Appendix,
$$
\frac{\partial }{\partial \rho} C(u_1,u_2) =\phi_2(z_1,z_2) .
$$
Here, $\phi_2$ is a bivariate normal probability density function and $z_j = \Phi^{-1}(u_j),$ $j=1,2$ are the normal scores corresponding to residuals $u_j$.

\subsection{Comparing Pairwise Likelihood to GMM Estimators}\label{S:Comparing}

This section compares the efficiency of the pairwise likelihood estimator $\theta_{PL}$ to the \textit{GMM} estimator $\theta_{GMM}$ through large sample approximations and (small sample) simulations.

\textbf{Asymptotic Comparison}. For large sample central limit theorem approximations, we have already stated that the asymptotic variance of $\theta_{GMM}$ is $\frac{1}{n} \left( G_{\theta}^{\prime} \left( \mathrm{Var~} g_{\theta} \right)^{-1} G_{\theta} \right)^{-1} .$

For a comparable statement for pairwise likelihoods, first note that the estimating equation can be expressed as
$$
g_{PL}(\mathbf{Y}_i; \theta) = \sum_{j=1}^{p-1} \sum_{k=j+1}^p  g_{\theta,ijk}(Y_{ij},Y_{ik})  =
\left(\mathbf{I}_r \otimes \mathbf{1}^{\prime}\right) g_{\theta,i}(\mathbf{Y}_i) = \mathbf{B~} g_{\theta,i}(\mathbf{Y}_i),
$$ where $\otimes$ is a Kronecker product, $\mathbf{I}_r$ is an $r \times r$ identity matrix, and $\mathbf{1}^{\prime}$ is a $1 \times \left(^p_2\right)$ vector of ones. From this, the sensitivity matrix is $\mathbf{B~} G_{\theta}$, where $G_{\theta}= \mathrm{E~} \frac{\partial}{\partial \theta^{\prime}} g_{\theta}$. The variance matrix is $ \mathbf{B~}\left( \mathrm{Var~} g_{\theta} \right)^{-1} \mathbf{B}^{\prime}$. Following the usual estimating equation methodology (c.f. \cite{song2007correlated}), the asymptotic variance of $\theta_{PL}$ is
$$\frac{1}{n} \left( G_{\theta}^{\prime} \mathbf{B}^{\prime} \left(\mathbf{B~} \mathrm{Var~} g_{\theta} ~ \mathbf{B}^{\prime}\right)^{-1} \mathbf{B~} G_{\theta} \right)^{-1} .$$
Not surprisingly, this is larger (in a matrix sense) than the \textit{GMM} estimator.

\textbf{Small Sample Comparison}. We also compared the efficiency of the pairwise likelihood and the \textit{GMM} estimator in a simulation study, the results are summarized in \textit{Online Supplement} 3. Not surprisingly, here we show that the \textit{GMM} estimator has a lower standard error than the pairwise estimator and in this sense is more efficient.

\section{Multivariate Longitudinal Modeling with Dropout}\label{S:LapseModel}

We now extend the \textit{GMM} estimation of copula regression models to more complex situations where the dependent outcomes include the observation process. This extension is motivated by the lapsation of insurance contracts where insurance customer lapsation is equivalent to biomedical patient dropout.

\subsection{Joint Model Specification}

Consider the case where we follow individuals (insurance policyholders) over time. For illustration, suppose that there are three outcome variables of interest. Two random variables, $Y_{1,it}$ and $Y_{2,it}$, might represent claims from auto and homeowners coverages, respectively. A third random variable, $L_{it}$, is a binary variable that represents a policyholder's decision to lapse the policy. Specifically, $L_{it}=1$ represents the policyholder's decision to not renew the policy (to lapse) and so we do not observe the policy at time $t+1$. If $L_{it}=0$, then we observe the policy at time $t+1$, subject to limitations on the number of time periods available. Let $m$ represent the maximal number of observations over time.

For each policyholder $i$ at time $t$, we potentially observe the lapse variable $L_{it}$ and a collection of other outcome variables $\mathbf{Y}_{it} = \{Y_{1,it}, \ldots, Y_{p,it} \}.$ Associated with each policyholder is a set of (possibly time varying) rating variables $\mathbf{x}_{it}$ for the $i$th policyholder at time $t$. Throughout, we assume independence among individuals.

To define the joint distribution among outcomes, we use copulas that are based on transformed outcomes. Specifically, for each $ijt$, define $\nu_{ijt} = \Phi^{-1}\left(F_{Y_{ijt}}(Y_{ijt})\right)$ to be the transformed outcome. In the same way, define $\nu_{iLt} = \Phi^{-1}\left(F_{L_{it}}(L_{it})\right)$. When the outcomes are continuous, the transformed outcomes have a normal distribution. For outcomes that are not continuous, one can utilize the generalized distributional transform of \cite{ruschendorf2009distributional}. With this generalization of the probability integral transform, it is straight-forward to show that a copula exists even when outcomes are not continuous.

\subsubsection{Temporal Structure}

Although copulas allow for a broad variety of dependence structures, for our longitudinal/panel data it is useful to employ an association structure motivated by traditional time series models as described in the following. In Section \ref{S:LapseIndependence}, we focus on the special case of no temporal dependence.

There are many ways of specifying simple parametric structures to capture the dependencies among these transformed outcomes. To capture dependencies, we use a standard multivariate time series framework due to Box and Jenkins (cf., \cite{tiao1981modeling}). In this framework, there are sequences of latent variables $\{\eta_{ijt}\}$ that are $iid$. Transformed outcomes can be expressed as
$ \nu_{ijt}= \eta_{ijt} +  \psi_{j,1} \eta_{ij,t-1} + \cdots +  \psi_{j,t-1} \eta_{ij,1}.$ For example, we might use a moving average of order one ($MA1$), $ \nu_{ijt}= \eta_{ijt} +  \psi_{j,1} \eta_{ij,t-1}$, or an autoregressive of order one ($AR1$) representation,
$ \nu_{ijt}= \eta_{ijt} +  \psi_{j,1} \nu_{ij,t-1} =  \eta_{ijt} +  \psi_{j,1} \eta_{ij,t-1} + \cdots +  \psi_{j,1}^{t-1} \eta_{ij,1}.$
Using matrix notation, we define $\boldsymbol \nu_{ij}= \left(\nu_{ij1}, \ldots, \nu_{ijm}\right)^{\prime}$, $\boldsymbol \eta_{ij}= \left(\eta_{ij1}, \ldots, \eta_{ijm}\right)^{\prime}$, and
$$\boldsymbol \Psi_j =\left(
\begin{array}{cccc}
1 & 0 & \cdots & 0 \\
\psi_{j,1} & 1 & \cdots & 0 \\
\vdots & \vdots && \vdots \\
\psi_{j,m-1} & \psi_{j,m-2} &  \cdots &1 \\
\end{array} \right) .$$
Thus, $\boldsymbol \nu_{ij} = \boldsymbol \Psi_j \boldsymbol \eta_{ij}$ and $\mathrm{Var~} \boldsymbol \nu_{ij} = \boldsymbol \Psi_j \boldsymbol \Psi_j^{\prime}$, for $j=1,\ldots, p$. In the same way, for lapse we have $\boldsymbol \nu_{iL} = \boldsymbol \Psi_L \boldsymbol \eta_{iL}$ and $\mathrm{Var~} \boldsymbol \nu_{iL} = \boldsymbol \Psi_L \boldsymbol \Psi_L^{\prime}$.

One can think of the $\eta_{ijt}$ as a transformed outcome with the temporal dependence filtered out. We further allow contemporaneous correlations among outcomes of the form
\begin{eqnarray*}
\mathrm{Cov}(\eta_{ijs}, \eta_{ikt})
 = \left\{
\begin{array}{cc}
\rho_{jk} & s=t \\
0 & s \ne t
\end{array} \right.
 \ \ \ \ \ \text{and} \ \ \ \ \
\mathrm{Cov}(\eta_{ijs}, \eta_{iLt})
 = \left\{
\begin{array}{cc}
\rho_{jL} & s=t \\
0 & s \ne t
\end{array} \right. .
\end{eqnarray*}
This yields
$
\mathrm{Cov}(\boldsymbol \nu_{ij}, \boldsymbol \nu_{ik}) = \rho_{jk} \boldsymbol \Psi_j \boldsymbol \Psi_k^{\prime}$  and
$\mathrm{Cov}(\boldsymbol \nu_{iL}, \boldsymbol \nu_{ij}) = \rho_{Lj} \boldsymbol \Psi_L \boldsymbol \Psi_j^{\prime} .$ We summarize this using the association matrix
\begin{equation}\label{E:Association1}
\bsS = \left(
\begin{array}{cccc}
\boldsymbol \Psi_L \boldsymbol \Psi_L^{\prime}    & \rho_{L1} \boldsymbol \Psi_L \boldsymbol \Psi_1^{\prime} & \cdots & \rho_{Lp} \boldsymbol \Psi_L \boldsymbol \Psi_p^{\prime} \\
\rho_{L1} \boldsymbol \Psi_1 \boldsymbol \Psi_L^{\prime}    & \boldsymbol \Psi_1 \boldsymbol \Psi_1^{\prime} & \cdots & \rho_{1p} \boldsymbol \Psi_1 \boldsymbol \Psi_p^{\prime} \\
\vdots & \vdots &\ddots  &\vdots \\
\rho_{Lp} \boldsymbol \Psi_p \boldsymbol \Psi_L^{\prime}    & \rho_{1p} \boldsymbol \Psi_p \boldsymbol \Psi_1^{\prime}& \cdots & \boldsymbol \Psi_p \boldsymbol \Psi_p^{\prime}
\end{array} \right) .
\end{equation}

To develop intuition, in Section \ref{S:LapseIndependence} we refer to the case where $\boldsymbol \Psi_j, j=1,\ldots,p,$ and $\boldsymbol \Psi_L$ have elements that are zero off the diagonal, the case of no temporal dependence.

\subsubsection{Joint Model}

To use the equation \eqref{E:Association1} association matrices, we restrict our attention to elliptical copulas and focus applications on the Gaussian copula. Extensions to other types of elliptical copulas (e.g., $t$-copulas) follow naturally.

It is convenient to introduce $T_i$, a variable that represents the time that the $i$th policyholder lapses. Specifically,
$$T_i = \left\{
\begin{array}{cl}
1 & \text{if }L_{i1}=1 \\
2 & \text{if }L_{i1}=0,L_{i2}=1 \\
\vdots & \ \ \ \ \ \vdots \\
t & \text{if }L_{i1}=0,\ldots,L_{i,t-1}=0, L_{i,t}=1 \\
\vdots & \ \ \ \ \  \vdots \\
m & \text{if }L_{i1}=0,\ldots,L_{i,m-1}=0, L_{im}=1 \\
m+1 & \text{if }L_{i1}=0,\ldots,L_{i,m}=0 .\\
    \end{array}
\right.
$$
 Now, suppose that we observe $T_i=t$ outcome periods. We will condition on the event $\{T_i=t\}$ that has probability
\begin{eqnarray}\label{E:Report1}
&&\Pr(T_i=t) = \Pr\left(L_{i1}=0,\ldots,L_{i,t-1}=0, L_{i,t}=1 \right)  \\
&& \ \ \ \ =
\left\{
\begin{array}{cc}
C\left(F_{Li,1}(0), \ldots, F_{Li,m}(0) \right) & t = m+1 \\
C\left(F_{Li,1}(0), \ldots, F_{Li,t-1}(0), 1, \ldots, 1 \right)-C\left(F_{Li,1}(0), \ldots, F_{Li,t}(0), 1, \ldots, 1 \right)& 1\le t \le m
    \end{array}
\right.  \notag \\
&& \ \ \ \ =
\left\{
\begin{array}{cc}
C\left(\mathbf{a}_{im}(F_{Li,m}(0))\right) & t = m+1 \\
C\left(\mathbf{a}_{it}(1)\right)-C\left(\mathbf{a}_{it}(F_{Li,t}(0)) \right)& 1 \le t \le m
    \end{array}
\right. . \notag
\end{eqnarray}
The last equality uses some notation $\mathbf{a}_{it}(a) = \left(F_{Li,1}(0), \ldots, F_{Li,t-1}(0),a, 1, \ldots, 1 \right)$ that we introduce to simplify the presentation. Calculation of these joint probabilities are straightforward, although tedious, from the marginals and the copula.

If a policy is renewed for all $m$ periods, then $T_i=m+1$ and the observed likelihood is based on
\begin{eqnarray}\label{E:Observed2}
&& \Pr \left(T_i =m+1, \mathbf{Y}_{i1} \le \mathbf{y}_{1}, \ldots, \mathbf{Y}_{im} \le \mathbf{y}_{m} \right) \\
&& \ \ \ \ =  \Pr \left(L_{i1}=0,\ldots,L_{i,m}=0, \mathbf{Y}_{i1} \le \mathbf{y}_1, \ldots, \mathbf{Y}_{im} \le \mathbf{y}_{m} \right) . \notag
\end{eqnarray}

If lapse occurs, then the observed likelihood is based on the distribution function
\begin{eqnarray}\label{E:Observed1}
&&\Pr \left(T_i=t, \mathbf{Y}_{i1} \le \mathbf{y}_{1}, \ldots, \mathbf{Y}_{it} \le \mathbf{y}_{t} \right) \\
&& \ \ \ \ = \Pr \left(
L_{i1}=0,\ldots,L_{i,t-1}=0, L_{i,t}=1, L_{i,t+1} \le \infty, \ldots, L_{im} \le \infty, \right.  \notag\\
&& \ \ \ \ \ \ \ \ \ \ \ \ \ \ \ \left. \mathbf{Y}_{i1} \le \mathbf{y}_{1}, \ldots, \mathbf{Y}_{it} \le \mathbf{y}_{t},\mathbf{Y}_{i,t+1} \le \infty, \ldots, \mathbf{Y}_{im} \le \infty \right) . \notag
\end{eqnarray}

Note that the evaluation of this likelihood involves a $m(p+1)$ dimensional copula. There are applications where this is computationally prohibitive and so standard maximum likelihood estimation is not available.

\subsection{Conditional Likelihood}\label{S:CondLikelihood}

Because of the difficulties in computing the observed likelihoods in equations \eqref{E:Observed1} and \eqref{E:Observed2}, we focus on the conditional distribution
\begin{eqnarray}\label{E:Conditional1}
 \Pr \left(Y_{j,is} \le y_1, Y_{k,is} \le y_2| T_i=t \right) &=&
 \frac{\Pr \left(T_i=t , Y_{j,is} \le y_1, Y_{k,is} \le y_2\right) }
 {\Pr \left(T_i=t \right) } .
\end{eqnarray}
Here, the time point $s$ is chosen so that $s \le \min(t,m)$. The subscripts $\{j,k\}$ represent any pair chosen from $\{1, \ldots, p\}$. We have already discussed the computation of the denominator in equation \eqref{E:Report1}. Computation of the numerator is similar; for the general case, it requires evaluation of a $m+2$ dimensional copula. As in equation \eqref{E:Report1},
\begin{eqnarray}\label{E:Conditional2}
&&\Pr(T_i=t, Y_{j,is} \le y_1, Y_{k,is} \le y_2)  \\
&& \ \ \ \ =
\left\{
\begin{array}{cc}
C\left(\mathbf{a}_{im}(F_{Li,m}(0)),F_{j,is}(y_1), F_{k,is}(y_2)\right) & s<t \le m+1 \\
C\left(\mathbf{a}_{it}(1),F_{j,is}(y_1), F_{k,is}(y_2)\right)-C\left(\mathbf{a}_{it}(F_{Li,t}(0)),F_{j,is}(y_1), F_{k,is}(y_2) \right)& s = t \le m
    \end{array}
\right. . \notag
\end{eqnarray}

We again remark that the copula $C$ in display \eqref{E:Conditional2} depends on the variables $j$ and $k$ selected as well as time points $s$ and $t$. Specifically, we use equation \eqref{E:Association1} to express the association matrix for $\{L_{i1}, \ldots, L_{i,m}, Y_{j,is}, Y_{k,is}\}$ as
\begin{equation}\label{E:Association2}
\bsS = \left(
\begin{array}{cccc}
\boldsymbol \Psi_L \boldsymbol \Psi_L^{\prime}               & \rho_{Lj} \boldsymbol \Psi_L \boldsymbol \Psi_{j,s}^{\prime} &  \rho_{Lk} \boldsymbol \Psi_L \boldsymbol \Psi_{k,s}^{\prime} \\
\rho_{Lj} \boldsymbol \Psi_{j,s} \boldsymbol \Psi_L^{\prime} & 1                                                        &  \rho_{jk} \boldsymbol \Psi_{j,s} \boldsymbol \Psi_{k,s}^{\prime} \\
\rho_{Lk} \boldsymbol \Psi_{k,s} \boldsymbol \Psi_L^{\prime}     & \rho_{jk} \boldsymbol \Psi_{j,s} \boldsymbol \Psi_{k,s}^{\prime}& 1
\end{array} \right) ,
\end{equation}
where $\boldsymbol \Psi_{j,s}$ is the $j$th row of $\boldsymbol \Psi_{j}$ and similarly for $\boldsymbol \Psi_{k}$.
\bigskip

\end{spacing}
The corresponding conditional hybrid probability density/mass functions follow as in Section \ref{S:PairwiseDist}. For $s<t \le m+1$, this is
\scalefont{.9}
\begin{equation}\label{E:ConditionalDensity1}
\begin{array}{ll}
f_{ijk,s|m+1}(y_1,y_2) = \frac{1}{\Pr(T_i=m+1)}  \times \\
\left\{
\begin{array}{ll}
\sum_{i_1=0}^1 \sum_{i_2=0}^1 \left(-1\right)^{i_1+i_2} C \left(\mathbf{a}_{im}(F_{Li,m}(0)),  F_{j,is}(y_1^{(i_1)}),F_{k,is}(y_2^{(i_2)})\right)
& y_1=y_1^d,y_2=y_2^d \\
\left\{ \sum_{i_2=0}^1 \left(-1\right)^{i_2}
C_{m+1}\left(\mathbf{a}_{im}(F_{Li,m}(0)), F_{j,is}(y_1), F_{k,is}(y_2^{(i_2)}) \right) \right\}f_{j,is}(y_1)
& y_1=y_1^c,y_2=y_2^d \\
\left\{  \sum_{i_1=0}^1 \left(-1\right)^{i_1} C_{m+2}\left(\mathbf{a}_{im}(F_{Li,m}(0)), F_{j,is}(y_1^{(i_1)}), F_{k,is}(y_2) \right)\right\} f_{k,is}(y_2)
& y_1=y_1^d,y_2=y_2^c \\
C_{m+1,m+2}\left(\mathbf{a}_{im}(F_{Li,m}(0)), F_{j,is}(y_1), F_{k,is}(y_2) \right)  f_{j,is}(y_1) f_{k,is}(y_2)
& y_1=y_1^c,y_2=y_2^c
\end{array} \right.
\end{array}
\end{equation}
\scalefont{1.1111}\begin{spacing}{1.8}

\noindent Here, we use the notation introduced in equation \eqref{E:MainLikelihood} where $y^{(i)} = y$ if $i=0$ and $y^{(i)} = y-$ if $i=1$. Further, $C_{m+1}(u_1,\ldots,u_m,u_{m+1},u_{m+2}) = \frac{\partial}{\partial u_{m+1}} C(u_1,\ldots,u_m,u_{m+1},u_{m+2})$ represents the partial derivative of the copula with respect to the second argument and similarly for $C_{m+2}$. The term $C_{m+1,m+2}$ is a second derivative with respect to the $m+1$st and $m+2$nd arguments. Further, $f_{j,is}$ is the density function corresponding to the distribution function $F_{j,is}$.

The corresponding conditional hybrid probability density/mass functions for $s = t \le m$, follows in the same way

%\scalefont{.9}
\begin{equation}\label{E:ConditionalDensity2}
\begin{array}{ll}
f_{ijk,s|t}(y_1,y_2) = \frac{1}{\Pr(T_i=t)}  \times \\
\left\{
\begin{array}{ll}
\sum_{i_r=0}^1 \sum_{i_1=0}^1 \sum_{i_2=0}^1 \left(-1\right)^{i_r+i_1+i_2}
\left\{
C \left(\mathbf{a}_{it}(F_{Li,t}(0)^{i_r}),  F_{j,is}(y_1^{(i_1)}),F_{k,is}(y_2^{(i_2)})\right)\right\}
& y_1=y_1^d,y_2=y_2^d \\
\left\{ \sum_{i_r=0}^1 \sum_{i_2=0}^1 \left(-1\right)^{i_r+i_2}
\left\{
C_{m+1}\left(\mathbf{a}_{it}(F_{Li,t}(0)^{i_r}), F_{j,is}(y_1), F_{k,is}(y_2^{(i_2)}) \right) \right\}
\right\}f_{j,is}(y_1)
& y_1=y_1^c,y_2=y_2^d \\
\left\{\sum_{i_r=0}^1  \sum_{i_1=0}^1 \left(-1\right)^{i_r+i_1}
\left\{
C_{m+2}\left(\mathbf{a}_{it}(F_{Li,t}(0)^{i_r}), F_{j,is}(y_1^{(i_1)}), F_{k,is}(y_2) \right)\right\}
\right\} f_{k,is}(y_2)
& y_1=y_1^d,y_2=y_2^c \\
\left\{\sum_{i_r=0}^1  \left(-1\right)^{i_r}
C_{m+1,m+2}\left(\mathbf{a}_{it}(F_{Li,t}(0)^{i_r}), F_{j,is}(y_1), F_{k,is}(y_2) \right)
\right\}
f_{j,is}(y_1) f_{k,is}(y_2)
& y_1=y_1^c,y_2=y_2^c
\end{array} \right.
\end{array}
\end{equation}
%\scalefont{1.1111}

\subsection{Dropout GMM Procedure}

We are now in a position to extend the generalized method of moments, GMM, procedure introduced earlier to incorporate lapse. As before, let $\theta$ be an $r$-dimensional vector that represents the parameters that quantify the association among $\{L_{it}, Y_{1,it}, \ldots, Y_{p,it} \}.$ Given $T_i=t$, the  hybrid probability density/mass function of $Y_{j,it}$ and $Y_{k,it}$ is $f_{ijk,s|t}(\cdot,\cdot)$, as specified in equations \eqref{E:ConditionalDensity1} and \eqref{E:ConditionalDensity2}.

To estimate $\theta$, for $s \le t$, define
\begin{eqnarray}\label{E:Score1}
g_{\theta,i,s,t}(y_1, y_2,T) = \mathrm{I}(T=t) ~ \partial_{\theta}  \ln f_{ijk,s|t}(y_1, y_2)   .
\end{eqnarray}
This is a mean zero random variable that contains information about $\theta$. To see that it has mean zero,
\begin{eqnarray*}
\mathrm{E~}g_{\theta,i,s,t}(Y_{j,is}, Y_{k,is},T_i)
&=& \mathrm{E~} \left[ \mathrm{E~} (g_{\theta,i,s,t}(Y_{j,is}, Y_{k,is},T_i) | T_i=t)\right] \\
&=& \mathrm{E~}  \left[ \mathrm{I}(T_i=t) \mathrm{E~} (\partial_{\theta}  \ln f_{ijk,s|t}(Y_{j,is}, Y_{k,is}) )\right] \\
&=& \mathrm{E~}  \left[ \mathrm{I}(T_i=t) \cdot 0 )\right] =0 ,
\end{eqnarray*}
using properties of a likelihood. Thus, $g_{\theta,i,s,t}(Y_{j,is}, Y_{k,is},T_i)$ is an unbiased estimator (of zero) and can be used in an estimating equation.

With the score function in equation \eqref{E:Score1}, we can use the \textit{GMM} procedure defined in Section \ref{S:GMMProcedure}. The only thing that remains to do is to evaluate the score functions in terms of copula-based functions.

\subsubsection{Dropout Score Evaluation}

The dropout scores are similar to the scores introduced in Section \ref{S:GMMScoreEval} except now we have a $m+2$ dimensional copula to evaluate instead of a 2-dimensional one. To provide additional intuition, we restrict ourselves to two random variables $Y_1$ and $Y_2$ that both follow a Tweedie distribution. These are non-negative random variables with continuous densities on the positive axis with a mass point at zero.

Thus, for $s<t \le m+1$ and two zero outcomes, this can be expressed as
\begin{eqnarray*}
 \partial_{\theta}  \ln f_{i12,s|m+1}(0,0) =
 \frac{\partial_{\theta} ~ C\left(\mathbf{a}_{im}(F_{Li,m}(0)), F_{1,is}(0), F_{2,is}(0) \right)}
{C\left(\mathbf{a}_{im}(F_{Li,m}(0)), F_{1,is}(0), F_{2,is}(0) \right)} -
 \frac{\partial_{\theta} ~ C\left(\mathbf{a}_{im}(F_{Li,m}(0)) \right)}
{C\left(\mathbf{a}_{im}(F_{Li,m}(0)) \right)}.
\end{eqnarray*}
For a single positive outcome, $y>0$, we have
\begin{eqnarray*}
\partial_{\theta}  \ln f_{i12,s|m+1}(y,0)
= \frac{\partial_{\theta} C_{m+1}\left(\mathbf{a}_{im}(F_{Li,m}(0)), F_{1,is}(y), F_{2,is}(0)\right)}
{C_{m+1}\left(\mathbf{a}_{im}(F_{Li,m}(0)), F_{1,is}(y), F_{2,is}(0)\right)}
 -
 \frac{\partial_{\theta} ~ C\left(\mathbf{a}_{im}(F_{Li,m}(0)) \right)}
{C\left(\mathbf{a}_{im}(F_{Li,m}(0)) \right)}
\end{eqnarray*}
For two positive outcomes, $y_1>0$ and $y_2>0$, we have
\begin{eqnarray*}
\partial_{\theta}  \ln f_{i12,s|m+1}(y_1,y_2)
= \frac{\partial_{\theta}  ~ C_{m+1,m+2}\left(\mathbf{a}_{im}(F_{Li,m}(0)), F_{1,is}(y_1), F_{2,is}(y_2) \right)}
                         {C_{m+1,m+2}\left(\mathbf{a}_{im}(F_{Li,m}(0)), F_{1,is}(y_1), F_{2,is}(y_2) \right)} -
 \frac{\partial_{\theta} ~ C\left(\mathbf{a}_{im}(F_{Li,m}(0)) \right)}
{C\left(\mathbf{a}_{im}(F_{Li,m}(0)) \right)} .
\end{eqnarray*}
Scores for $s \le t = m$ follow in a similar way. In \textit{Online Supplement} 2, we give explicit formulas for Gaussian copula derivatives.

\section{Dropout Modeling with Temporal Independence}\label{S:LapseIndependence}

In this section, we assume no temporal dependence for dropout and outcomes. The purpose is to provide a context that is still helpful in applications and allows us to develop more intuition. To aid with developing intuition, we focus on the case where $p=2$. Even though there is no temporal dependence, we can use a copula function to capture dependence among dropout and outcomes. This specification permits, for example in our insurance setting, large claims to influence the tendency to lapse a policy or a latent variable to simultaneously influence both lapse and claims outcomes.

\subsection{Joint Model Specification}\label{S:JointModelNoTimeDepend}

With no temporal dependence, we assume that the random variables in $\{L_{it}, \mathbf{Y}_{it}\}$ are independent over $i$ and $t$. With the independence over time, the joint distribution function of the potentially observed variables is
\begin{eqnarray*}
&&\Pr \left(L_{i1} \le r_1, \ldots, L_{im} \le r_m, \mathbf{Y}_{i1} \le \mathbf{y}_{1}, \ldots, \mathbf{Y}_{im} \le \mathbf{y}_{m} \right)\\
&& \ \ \ \ = \prod_{t=1}^m C \left(F_{Lit}(r_t), F_{1,it}(y_{1t}), F_{2,it}(y_{2t}) \right).
\end{eqnarray*}
As before, $F_{Lit}$ and $F_{j,it}$ represent the marginal distributions of $L_{it}$ and $Y_{j,it}$,  $j=1, 2$.

In the case of temporal independence, display \eqref{E:Report1} reduces to
\begin{eqnarray*}\label{E:Report3}
\Pr(T_i=t)  &=&
\left\{
\begin{array}{cc}
\prod_{s=1}^m F_{Li,s}(0) & t=  m+1  \\
\left(1-F_{Li,t}(0) \right)\prod_{s=1}^{t-1} F_{Li,s}(0) & 1 \le t \le m
    \end{array}
\right. .
\end{eqnarray*}

From this, we will be able to use dropouts to estimate the marginal distribution in the usual fashion. Intuitively, this is because the claims outcomes do not affect our ability to observe lapse outcomes. The converse is not true, lapses do affect our ability to observe claims. However, this is not true of first period claims. These are always observed in this model and so provide the basis for consistent estimates of the claims marginal distributions.

If a policy is renewed for all $m$ periods, then $T_i=m+1$ and with \eqref{E:Observed2}, the observed likelihood is based on
\begin{eqnarray}\label{E:Observed4}
\Pr \left(T_i =m+1, \mathbf{Y}_{i1} \le \mathbf{y}_{1}, \ldots, \mathbf{Y}_{im} \le \mathbf{y}_{m} \right) = \prod_{s=1}^{m} C\left(F_{Lis}(0) , F_{1,is}(y_{1s}), F_{2,is}(y_{2s}) \right) . \notag
\end{eqnarray}

If lapse occurs, then $T_i \le m$ and with \eqref{E:Observed1}, the observed likelihood is based on the distribution function
\begin{eqnarray}\label{E:Observed3}
&&\Pr \left(T_i=t, \mathbf{Y}_{i1} \le \mathbf{y}_{1}, \ldots, \mathbf{Y}_{it} \le \mathbf{y}_{t} \right) \notag\\
&& \ \ \ \ = \left\{C\left(1, F_{1,it}(y_{1t}),F_{2,it}(y_{2t}) \right)-C\left(F_{Lit}(0) , F_{1,it}(y_{1t}),F_{2,it}(y_{2t}) \right)\right\} \notag  \\
&& \ \ \ \ \ \ \ \ \ \ \ \ \
\prod_{s=1}^{t-1} C\left(F_{Lis}(0) , F_{1,is}(y_{1s}),F_{2,is}(y_{2s}) \right) . \notag
\end{eqnarray}

Thus, the corresponding conditional distribution function is
\begin{eqnarray}\label{E:Observed5}
&&\Pr \left(Y_{1,is} \le y_1, Y_{2,is} \le y_2 |T_i=t \right) = \\
&& \ \ \ \ \ \ \ \ \ \ \ \ \ \ \left\{\begin{array}{cl}
\frac{C\left(F_{Lis}(0), F_{1,is}(y_1), F_{2,is}(y_2) \right)}{F_{Lis}(0)} & \text{for  } s<t \le m+1 \\
\frac{C\left(1, F_{1,is}(y_{1}),F_{2,is}(y_{2}) \right)-C\left(F_{Lis}(0) , F_{1,is}(y_{1}),F_{2,is}(y_{2}) \right)}
{1-F_{Lis}(0)}  & \text{for  }s = t \le m
\end{array}\right. . \notag
\end{eqnarray}
This is more intuitive than the general expressions given in Section \ref{S:CondLikelihood}. Note that the evaluation of this distribution function involves a $3$ dimensional copula.

Expressions for the conditional density/mass functions and the \textit{GMM} scores follow a similar pattern and are omitted here.

\subsection{Missing at Random}\label{S:MAR}

The no temporal dependence assumption provides one additional benefit; the decision to lapse turns out to be \textit{ignorable} in the sense that the response mechanism does not affect inference about claims. This is in spite of the fact that we still allow for a dependency between claims and response; we note that this is not consistent with the usual biomedical literature on joint models of longitudinal and time-to-event data, c.f., \cite{elashoff2017joint}. Intuitively, this is because, in the insurance setting, we postulate a dependency between a claim during the $t$th year, $Y_t$, and a decision lapse or renew \textbf{at} time $t$, $L_t$. In contrast, in biomedical applications, the concern is for dependencies between whether an outcome (claims) is observed and a variable to indicate whether it is observed. In our notation, if $Y_t$ is the random variable in question, then $L_{t-1}$ indicates whether or not it is observed (lapse/renewal at time $t-1$).

More formally, consider a joint probability mass/density function of the form $Like^* = f(Y, L | \mathbf{X}, \boldsymbol \theta)$ where $\mathbf{X}$ are covariates and $\boldsymbol \theta$ are parameters. By conditioning, write this as $Like^* = \prod_{s=1}^m  f(Y_s, L_s | \mathbf{X}, \boldsymbol \theta, H_s)$ with history $H_s = \{Y_1, \ldots, Y_{s-1}, L_1, \ldots, L_{s-1} \}$. On the event $\{T = t\} = \{L_1=1, \ldots, L_{t-1}=0, L_t = 1, \ldots, L_m = 1 \}$, we have
\begin{eqnarray*}
Like^* &=& \left( \prod_{s=1}^{t-1}  f(Y_s, L_s=0 | \mathbf{X}, \boldsymbol \theta, H_s )\right)
 f(Y_t, L_t=1 | \mathbf{X}, \boldsymbol \theta, H_s )
\left( \prod_{s=t+1}^m  f(Y_s, L_s=1 | \mathbf{X}, \boldsymbol \theta, H_s )\right) \\
 &=&\left( \prod_{s=1}^{t-1}  f(Y_{obs,s}, L_s=0 | \mathbf{X}, \boldsymbol \theta, H_s )\right)
 f(Y_{obs,t},Y_{mis,t}, L_t=1 | \mathbf{X}, \boldsymbol \theta, H_s ) \\
&& \ \ \ \ \ \ \ \ \ \left( \prod_{s=t+1}^m  f(Y_{mis,s}, L_s=1 | \mathbf{X}, \boldsymbol \theta, H_s )\right),
\end{eqnarray*}
\noindent where $Y_{obs}, Y_{mis}$ signals that the claim is observed or missing (unobserved), respectively. In the last term, the event $\{L_s=1\}$ is known given $H_s$ due to the monotonicity of lapsation. Further, because of independence, the last term is
\begin{eqnarray*}
 \prod_{s=t+1}^m  f(Y_{mis,s}, L_s=1 | \mathbf{X}, \boldsymbol \theta, H_s ) =
\prod_{s=t+1}^m  f(Y_{mis,s}| \mathbf{X}, \boldsymbol \theta )
\end{eqnarray*}
which can be integrated out to get the (observed) likelihood. In our insurance sampling scheme, the middle term is
$ f(Y_{obs,t},Y_{mis,t}, L_t=1 | \mathbf{X}, \boldsymbol \theta, H_s ) =
 f(Y_{obs,t}, L_t=1 | \mathbf{X}, \boldsymbol \theta, H_s ) .$ Thus, the integrated likelihood is
\begin{eqnarray*}
Like
 &=&\left( \prod_{s=1}^{t-1}  f(Y_{obs,s}, L_s=0 | \mathbf{X}, \boldsymbol \theta, H_s )\right)
 f(Y_{obs,t} L_t=1 | \mathbf{X}, \boldsymbol \theta, H_s )
\end{eqnarray*}
which is the observed likelihood. This establishes our claim that the lapse decision is ignorable in the case of temporal independence.

\subsection{Comparing Trivariate Likelihood to GMM Estimators}\label{S:LapseComparing}

Similar to the simulation study cited at the end of Section \ref{S:Comparing}, we consider a sample of $n$ policyholders that are potentially observed over 5 years. In addition to the claims, we have five rating (explanatory) variables: (a) $x_1$, a binary variable that indicates whether or not an attribute holds, (b) $x_2$, $x_3$, $x_4$, are generic continuous explanatory variables, and (c) $x_5$ is a  time trend ($x_{it} = t$).

For the claims variables, we used a logarithmic link to form the mean claims $\mu_{j,it} = \exp\left(\mathbf{x}_{it}^{\prime} \boldsymbol \beta_j\right),$ $j=1,2$. For the lapse variable, the expected value is $\pi_{it} = \exp\left(\mathbf{x}_{it}^{\prime} \boldsymbol \beta_L\right)/\left(1+\exp\left(\mathbf{x}_{it}^{\prime} \boldsymbol \beta_L\right)\right),$ a common form for the logit model. We use a negative coefficient associated with the time trend variable to reflect the fact that lapse probabilities tend to decrease with policyholder duration.

Each type of claims was simulated using the Tweedie distribution, a mean, and two other parameters, $\phi_j$ (for dispersion) and $P_j$ (the ``power'' parameter). Recall, for a Tweedie distribution, that the variance is $\phi_j \mu^{P}$; we use $P=1.67$ in this study.

Dependence among claims was taken to be a Gaussian copula with the following structure
$$
\boldsymbol \Sigma  = \left(
\begin{array}{ccccc}
1         & \rho_{L1}  & \rho_{L2} \\
\rho_{L1} & 1          & \rho_{12} \\
\rho_{L2} & \rho_{12}  & 1 \\
    \end{array}
\right) .
$$
For example, we might use positive values for the association between lapse and claims ($\rho_{L1}$ and $\rho_{L2}$) so that large claims are associated with higher lapse. We might use a positive value for the association between claim types ($\rho_{12}$).

This section compares the efficiency of the pairwise likelihood estimator $\theta_{PL}$ to the \textit{GMM} estimator $\theta_{GMM}$.

Table \ref{T:CompareLapse} summarizes the performance of the \textit{GMM} estimators lapse estimator by varying the sample size and dispersion parameters, $\phi_1$, and $\phi_2$. For this table, we used $\rho_{L1} = -0.2$, $\rho_{L2} = 0.2$, and $\rho_{12} = 0.1$ for the association parameters, these being comparable to the results of our empirical work. For smaller samples, $n=100, 250$, we used 500 simulations to make sure that the bias was being determined appropriately. This was less of a concern with larger sample sizes, $n=500,1000, 2000$, and so for convenience we used 100 simulations in this study.

Some aspects of the results are consistent with our Table \ref{T:CompareLapse} which compares pairwise likelihood and \textit{GMM} estimators study (without lapse). As the dispersion parameters $\phi$ increase, there are more discrete observations resulting in larger biases and standard errors for all sample sizes. The magnitude of biases suggests that our general procedure may not be suitable for sample sizes as small as $n=100$. However, even for $n=250$ (and above), we deem their performance acceptable on the bias criterion.

For the standard error criterion, we view the smaller sample sizes $n=100,250$ as unacceptable. For example, if $n=100$, $\phi_1=\phi_2=500$, and $\rho_{L1} = -0.2$, it is hard to imagine recommending a procedure where the average standard error is 0.158. Only for nearly continuous data, when $\phi_1=\phi_2=2$, do the standard errors seem desirable with $n=500$. In general, for more discrete data where $\phi_1=\phi_2=500$, we recommend samples sizes of $n=2,000$ and more. Most users that we work with are primarily interested in point estimates but also want to say something about statistical significance.

We use the pairwise estimators as starting values in the more complete \textit{GMM} estimators. As documented in \textit{Online Supplement} 4, the \textit{GMM} estimators are marginally more efficient than the pairwise estimators. Moreover, we investigated the performance of the \textit{GMM} estimators by varying the association parameters and learned that their performance was robust to the choice of different sets of association parameters.

\end{spacing}
\begin{table}[!tbp]
\begin{center}
\caption{Summary of the \textit{GMM} Lapse Estimators}
 \scalefont{0.9}
\begin{tabular}{rrr|rrr|rrr}\hline
Num & $n$ & $\phi_1=\phi_2$ & \multicolumn{3}{c|}{Bias} & \multicolumn{3}{c}{Standard Error} \\
Sim &     &                 & $L1$ & $L2$ & $12$              & $L1$ & $L2$ & $12$ \\ \hline
    500   & 100   & 2     & 0.000 & 0.003 & -0.001 & 0.092 & 0.091 & 0.049 \\
    500   & 100   & 42    & -0.006 & -0.011 & -0.002 & 0.096 & 0.094 & 0.053 \\
    500   & 100   & 500   & -0.023 & -0.006 & -0.035 & 0.158 & 0.139 & 0.135 \\ \hline
    500   & 250   & 2     & -0.001 & -0.002 & 0.001 & 0.059 & 0.058 & 0.031 \\
    500   & 250   & 42    &  0.005 & 0.002 & 0.000 & 0.062 & 0.060 & 0.034 \\
    500   & 250   & 500   & -0.004 & -0.012 & -0.011 & 0.107 & 0.091 & 0.086 \\ \hline
    100   & 500   & 2     & -0.001 & 0.000 & -0.002 & 0.042 & 0.041 & 0.022 \\
    100   & 500   & 42    & -0.002 & -0.005 & -0.003 & 0.044 & 0.043 & 0.024 \\
    100   & 500   & 500   & -0.011 & -0.009 & -0.008 & 0.076 & 0.064 & 0.061 \\ \hline
    100   & 1000  & 2     & -0.002 & 0.002 & 0.000 & 0.030 & 0.029 & 0.015 \\
    100   & 1000  & 42    &  0.009 & 0.001 & -0.002 & 0.031 & 0.030 & 0.017 \\
    100   & 1000  & 500   & -0.002 & -0.003 & -0.003 & 0.054 & 0.045 & 0.044 \\ \hline
    100   & 2000  & 2     &  0.000 & 0.001 & 0.001 & 0.021 & 0.021 & 0.011 \\
    100   & 2000  & 42    &  0.004 & -0.005 & 0.001 & 0.022 & 0.021 & 0.012 \\
    100   & 2000  & 500   &  0.004 & 0.000 & -0.003 & 0.038 & 0.032 & 0.031 \\
\hline
\end{tabular}\label{T:CompareLapse}
\end{center}\scalefont{1.1111}
\end{table}\begin{spacing}{1.8}

\section{Empirical Application}\label{S:Empirical}

\subsubsection*{Insurance Lapsation}

Life insurers enter into contracts that can last many years (e.g., a 20 year old purchasing a policy who dies at age 100 has an 80 year contract). In contrast, property and casualty (also known as general and as property and liability) insurers write contracts of shorter durations, typically six months or a year. In both cases, the company anticipates policyholders will retain a relationship with the insurer for many years and they track when and why policyholders leave, or lapse. Although policyholders may depart at any time, lapsation tends to occur at a periodic premium payment times and so we follow standard industry practice and treat time as discrete.

Why do insurers worry about lapsation? They seek to (a) retain profitable customers for a longer period of time,
(b) achieve profit margins on new customers within short time, and (c) achieve higher profit margins on existing customers.
See, for example, \cite{guillen2003using}, \cite{brockett2008survival}, and \cite{guillen2012time}.

There is a natural struggle between the price of an insurance risk and loyalty. The higher the price, the higher is the probability to lapse the policy. Despite this relationship, insurers typically separate the processes of calculating price and renewal prospects. They use standard event-time models to understand characteristics of policyholders that drive renewal or lapsation.

\subsubsection*{Data and Variable Descriptions}

This paper examines longitudinal data from a major Spanish insurance company that offers automobile and homeowners insurance. As in many countries, in Spain vehicle owners are obliged to have some minimum form of insurance coverage for personal injury to third parties. Homeowners insurance, on the other hand, is optional. The dataset tracks 40,284 clients over five years, between 2010 and 2015, who subscribed to both automobile and homeowners insurance. From the unbalanced panel of policyholders over 5 years, there are $N=122,935$ observations in the data set. Table \ref{T:NumberLapse} summarizes lapse behavior. The Spanish market is competitive; the table shows 29,296 lapses, for a lapse rate of 23.8\%.

\end{spacing}
\begin{table}[!tbp]
\begin{center}
\caption{Number of Policies and Lapse by Year}
 \scalefont{0.9}
\begin{tabular}{lrrrrr}
\hline \hline
                           &2010& 2011 & 2012 & 2013 & 2014 \\
\hline
 Number of customers at \\
 the beginning of the period  & 40284 &	29818 	&22505 &	17044 &	13284\\
 Number of customers that  &  &  &  &  \\
 cancel at least one policy  &  &  &  &  \\
 per period  & 10466 &	7313 &	5461 	&3760 	&2296 \\
 Rate of lapsation for  &  &  &  &  \\
 at least one policy (\%)& 26\% &25\% & 	24\% & 	22\% & 	17\%  \\
\hline \hline
\end{tabular}\label{T:NumberLapse}
\scalefont{1.111}
\end{center}
\end{table}

\begin{spacing}{1.8}

Our database includes variables that are commonly used for determining prices and understanding lapse behavior. These include (a) customer characteristics such as age and gender, (b) vehicle characteristics for auto insurance, (c) information on property for homeowners, and (d) renewal information such as the date of renewal. Table \ref{T:VarDescribe} provide variable descriptions.

\end{spacing}
\begin{table}[!tbph]
\begin{center}
\caption{Variable Descriptions}
 \scalefont{0.9}
\begin{tabular}{l|l}
\hline \hline
Variable	&  Description \\
\hline
year                 & \\
gender               & 1 for male, 0 for female\\
Age\_client          &age of the customer\\
Client\_Seniority     &the number of years with the company\\
metro\_code             &1 for urban or metropolitan, 0 for rural\\
Car\_power\_M               &power of the car\\
Car\_2ndDriver\_M             &presence of a second driver\\
Policy\_PaymentMethodA        &1 for annual payment, 0 for monthly payment\\
Insuredcapital\_continent\_re &value of the property\\
appartment                    &1 for apartment, 0 for houses \\
                              & ~ ~ ~ ~ ~~~ or semi-attached houses\\
Policy\_PaymentMethodH        &1 for annual payment, 0 for monthly payment\\
\hline \hline
\end{tabular}\label{T:VarDescribe}
\scalefont{1.111}
\end{center}
\end{table}
\begin{spacing}{1.8}

Table \ref{T:ClaimSumStats} gives basic frequency and severity summary statistics for auto and homeowners claims.  For type 1 (auto) claims, we have 1,967 or about 1.6\% claims. For type 2 (home) claims, we have 2,189 or about 1.78\%  claims. Readers interested in details of this dataset may look to the \textit{Online Supplement} 5 which gives many more summary statistics.

\end{spacing}
\begin{table}[!tbph]
\begin{center}
\caption{Claim Summary Statistics}
 \scalefont{0.9}
\begin{tabular}{l|rrrrr}
\hline \hline
                         &2010  & 2011 & 2012 & 2013 & 2014 \\
\hline
Clients with positive claims  - Auto  &        769 &        547 &        318 &        209 &        124 \\
Average number of claims - Auto&       0.04 &       0.03 &       0.03 &       0.02 &       0.02 \\
Average claim amount - Auto (Euros) &    1539.99 &    1689.84 &     2031.2 &    1629.18 &    1222.13 \\
Clients with positive claims  - Home &        660 &        531 &        448 &        310 &        240 \\
Average number of claims  - Home &       0.03 &       0.03 &       0.04 &       0.03 &       0.03 \\
Average claim amount  - Home  (Euros) &     447.85 &     501.59 &     410.73 &      348.1 &     508.86 \\
\hline \hline
\end{tabular}\label{T:ClaimSumStats}
\scalefont{1.111}
\end{center}
\end{table}
\begin{spacing}{1.8}

\subsubsection*{Marginal Model Fits}

After extensive diagnostic testing described in the \textit{Online Supplement} 5, we fit standard generalized linear models to the three outcome variables, comparable to insurance industry practice. Specifically, we fit a logistic model for lapse and Tweedie regression models for auto and homeowners claims. Table \ref{T:MarginalModelFits} summarizes estimates of marginal model fits.

\end{spacing}
\begin{table}[!tbph]
\begin{center}
\caption{Marginal Models Fits}
 \scalefont{0.9}
\begin{tabular}{l|rr|rr|rr}
\hline \hline
 & \multicolumn{2}{c|}{Logistic -- Lapse} & \multicolumn{2}{c|}{Tweedie -- Auto}  & \multicolumn{2}{c}{Tweedie -- Home}\\
    {\bf } & {\bf Estimate} & {\bf t value} & {\bf Estimate} & {\bf t value} & {\bf Estimate} & {\bf t value} \\
    \hline
(Intercept) &      0.324 &       9.39 &     21.132 &       5.58 &     -2.794 &      -2.58 \\
      year &     -0.078 &     -15.19 &     -1.179 &      -2.68 &     -0.015 &      -0.32 \\
    gender &      0.097 &       5.89 &            &            &            &            \\
Age\_client &     -0.023 &     -41.18 &     -0.421 &      -7.59 &      0.013 &       2.60 \\
Client\_Seniority &     -0.006 &      -4.50 &      0.170 &       0.63 &     -0.004 &      -0.33 \\
metro\_code &      0.163 &       9.13 &     -4.356 &      -2.25 &      0.261 &       1.57 \\
Car\_power\_M &            &            &      0.122 &      23.54 &            &            \\
Car\_2ndDriver\_M &            &            &     -2.354 &      -0.49 &            &            \\
Policy\_PaymentMethodA &            &            &      3.542 &       2.68 &            &            \\
Insuredcapital\_continent\_re &            &            &            &            &      0.348 &       4.25 \\
appartment &            &            &            &            &      1.097 &       7.16 \\
Policy\_PaymentMethodH &            &            &            &            &     -0.765 &      -3.40 \\
\hline \hline
\end{tabular}\label{T:MarginalModelFits}
\scalefont{1.111}
\end{center}
\end{table}
\begin{spacing}{1.8}

\subsubsection*{Interpreting the Results}

Using the marginal fitted regression models as inputs, we then computed the generalized method of moments association parameters. The results are summarized in Table \ref{T:AssociationEstimates}.

\end{spacing}
\begin{table}[!tbph]
\begin{center}
\caption{GMM Association Estimates}
 \scalefont{0.9}
\begin{tabular}{l|rrrr}
\hline \hline
    {\bf } & {\bf Estimate} & {\bf Std Error} & {\bf t value} \\
\hline
Lapse-Auto &      0.101 &      0.007 &      14.14 \\
Lapse-Home &      0.069 &      0.011 &       6.11 \\
 Auto-Home &      0.118 &      0.029 &       4.10 \\
\hline \hline
\end{tabular}\label{T:AssociationEstimates}
\scalefont{1.111}
\end{center}
\end{table}
\begin{spacing}{1.8}

These results show that even after controlling for the effects of the characteristics of the client, vehicles and homes, there is still evidence of relationship between claims for auto insurance, claims for home insurance and the renewal behavior.

Specifically, for a customer with a claim, there is a higher tendency to lapse. From the consumers perspective, a claim may induce a search for a new company (possibly for fear of experience rating induced premium increases). With no claims, insureds may be content to remain loyal to a company (perhaps simply inertia). From the company perspective, a claim is an alert for non-renewal, something that the insurer can take action upon should they wish to.

We also note that the association between auto and home claims has implications for portfolio management and reserving practices -- these two risks are not independent.

Academic pricing models are anchored to the notion that personal lines is a short-term business because contracts typically are 6 months or a year. These pricing models are based on costs of insurance so from an economics perspective can be thought of as ``supply-side'' models. In contrast, insurers consider non-renewals a critical disruption for their business - losing a customer implies that they will not see benefits in renewal years. Although the contract is only for 6 months or a year, customers are likely to renew and this is factored into models of insurer profitability. In fact, some insurers explicitly seek to maximize what is known as ``lifetime customer value'' where renewal is a key feature. Because renewal is about who buys insurance, from an economics perspective it can be thought of as ``demand-side'' modeling. Our joint model of renewal and insurance risk may play a key role in developing models that balance demand and supply side considerations of personal insurance lines.

\section{Summary and Conclusions}\label{S:Conclusion}

Motivated by insurance applications, this paper has introduced the \textit{GMM} procedure to estimate association parameters in complex copula regression modeling situations where the marginals may be a hybrid combination of discrete and continuous components. Because the \textit{GMM} scores use only a low-dimensional subset of the data, this procedure is available in high-dimensional situations. Thus, it provides an alternative to the vine copula methodology, c.f., \cite{panagiotelis2012pair}. Compared to the \textit{GMM} approach, the advantage of vines is the flexibility in model specification where many different sub-copula models may be joined to represent the data. The comparative advantage of the \textit{GMM} approach may be the relative efficiency. For example, we have seen in Section \ref{S:CopRegression} that the \textit{GMM} approach is more efficient than the pairwise likelihood. A strength of the \textit{GMM} approach is that, in Section 2, it uses scores based on pairs of relationships and so is able to take advantage of technologies developed for paired copulas, e.g., \cite{schepsmeier2014derivatives}.

Moreover, the \textit{GMM} approach was extended in Sections \ref{S:LapseModel}-\ref{S:Empirical} to handle problems of data attrition. Here, one needs to condition on the observation process and so at least three observations are required to calibrate associations among claims. Section \ref{S:LapseModel} provides a detailed theory on handling general longitudinal data with explicit expressions in Section \ref{S:LapseIndependence} for temporally independent, yet still dependent censoring, of observations. The supporting Appendices extend some of the pairwise technologies to multiple dimensions and make connections with an earlier literature on sensitivity of association parameters based on work of \cite{plackett1954reduction}.

Insurance analysts employ regression techniques to address the heterogeneity of customers. Copulas provide an interpretable way of incorporating dependence, that is fundamental to insurance operations, while preserving marginal models of claims outcomes. As in many social science and biomedical fields where subjects are followed longitudinally, problems of attrition arise that, in the insurance context, are known as lapsation. In this paper, we argue that a copula approach that allows one to handle attrition, in addition to many other potential sources of dependence, is a natural modeling strategy for analysts to adopt.

\end{spacing}

\scalefont{0.8}
\bibliography{LapseJan2019}
\scalefont{1.25}

\section{Appendix. Derivatives with Respect to Association Parameters}\label{S:Appendix2}

\subsection{Bivariate Gaussian Copula}\label{S:BivariateGaussian}

For derivatives with respect to $\rho$, we start with the distribution function and note a result that, according to \cite{plackett1954reduction}, was well known even in the mid-1950's,
$$ \frac{\partial }{\partial \rho} \Phi_2(z_1,z_2) = \phi_2(z_1,z_2). $$
For Gaussian copulas, this immediately yields
$$ \frac{\partial }{\partial \rho} C(u_1,u_2) =\phi_2(z_1,z_2). $$

For other details in the bivariate case, we use the work of \cite{schepsmeier2012web, schepsmeier2014derivatives}. For example, they calculate the partial derivative
$$
\frac{\partial C_2(u_1,u_2)}{\partial \rho} =
\phi \left(\frac{\Phi^{-1}(u_1) - \rho \Phi^{-1}(u_2)}{\sqrt{1 - \rho^2}}\right) \cdot
\frac{\rho \Phi^{-1}(u_1) -\Phi^{-1}(u_2)}{(1 - \rho^2)^{3/2}} .
$$

\subsection{Classic Multivariate Result}
In the following, we sometimes require the correlation matrix associated with $\bsS$, defined as
$$ \mathbf{R} = \mathrm{R}(\bsS) = (\mathrm{diag} (\bsS))^{-1/2} ~\bsS ~(\mathrm{diag} (\bsS))^{-1/2} .$$
Further, following standard notation, the subscripts in $\mathbf{R}_{ij}$ refer to the $i$th row and $j$th column of the matrix $\mathbf{R}$.

For a general multivariate approach, we cite a result due to \cite{plackett1954reduction}. Partition the correlation matrix as
$$
 \mathbf{R}= \left(
\begin{array}{cc}
 \mathbf{R}_{1:2,1:2} &  \mathbf{R}_{1:2,3:d}\\
 \mathbf{R}_{1:2,3:d}^{\prime} &   \mathbf{R}_{3:d,3:d}
\end{array}
\right)
 \ \ \ \ \ \ \
\mathbf{R}_{1:2,1:2} = \left(
\begin{array}{cc}
1 & \rho\\
\rho &  1
\end{array}
\right) ,
$$
so that  $\mathbf{R}_{1:2,1:2}$ is the submatrix for the first two elements and $\rho$ is the corresponding correlation coefficient. Then, from \cite{plackett1954reduction}, we have
\begin{eqnarray}\label{E:hFunction2a}
h_{2,d}^{(12)}(x_1, \ldots, x_d; \bsS) &=& \frac{\partial} {\partial \rho} \Phi_d(x_1, \ldots, x_d; \mathbf{R})\\
 &=& \phi_2\left(x_1 , x_2 ; \mathbf{R}_{1:2,1:2}\right)
 \Phi_{d-2}(\mathbf{x}^*;  \mathbf{R}_{\{3:d,3:d\}\cdot \{1:2\}}) \notag,
\end{eqnarray}
where
$$
\mathbf{x}^*  =
\left(  \begin{array}{c} x_3 \\ \vdots \\ x_d  \end{array} \right)
-  \mathbf{R}_{1:2,3:d}^{\prime} \mathbf{R}_{1:2,1:2}^{-1}
\left(  \begin{array}{c} x_1  \\  x_2 \end{array} \right)
\ \ \ \text{and} \ \ \
\mathbf{R}_{\{3:d,3:d\}\cdot \{1:2\}} = \mathbf{R}_{3:d,3:d}
-  \mathbf{R}_{1:2,3:d}^{\prime} \mathbf{R}_{1:2,1:2}^{-1} \mathbf{R}_{1:2,3:d}  .
$$
In the same way, define $h_{2,d}^{(ij)}$ by interchanging the first and second random variables with the $(i,j)$ pair of random variables.

See also \cite{gassmann2003multivariate} for a more recent commentary on this interesting result.

\subsection{Association Parameter Derivatives}

\subsubsection{Association Parameter Derivatives of $C_d$}

We now seek to compute the partial derivative with respect association parameter of the partial copula function (the partial copula function is reviewed in \textit{Online Supplement} 2). This a function of $d-1$ arguments of the form $z_j - \mu_{1\cdot2,j}$ and an additional ${d \choose 2}$ arguments in $\mathbf{R}$. For the first set of arguments, we use the expression
\begin{eqnarray}\label{E:hFunction1}
h_{1,d}^{(1)}(x_1, \ldots, x_d; \mathbf{R}) &=& \frac{\partial} {\partial x_1} \Phi_d(x_1, \ldots, x_d; \mathbf{R}) \notag\\
&=& \phi\left(x_1\right)\Phi_{d-1}(x_2-x_1^*, \ldots, x_d-x_1^*; \mathbf{R}_{2 \cdot 1}) ,
\end{eqnarray}
where
$$ x_1^* = x_1^*(\mathbf{R}) =\mathbf{R}_{2:d,1} ~x_1
\ \ \ \text{and} \ \ \
 \mathbf{R}_{2 \cdot 1} = \mathbf{R}_{2:d,2:d} - \mathbf{R}_{2:d,1}  \mathbf{R}_{2:d,1}^{\prime} . $$
We define $h_{1,d}^{(j)}$ in the same way by interchanging the first and the $j$th random variables.

The relationship in equation \eqref{E:hFunction1} can be easily established using two facts. The first is that, for a general multivariate distribution, we have
$$\frac{\partial}{\partial x_1} \Pr(X_1\le x_1, X_2\le x_2, \ldots, X_d\le x_d) = f_{X_1}(x_1) \Pr( X_2\le x_2 \ldots, X_d\le x_d| X_1 = x_1) .$$
The second is that, for $X_1, \ldots, X_d$ is normal with mean zero and variance-covariance matrix $\mathbf{R}$, we have $X_2, \ldots, X_d$ given $X_1$ is normal with mean $ x_1^*$ and variance-covariance $ \mathbf{R}_{2 \cdot 1}$.

Define the rescaled arguments \newline$z_j^* = \left(z_j - \mu_{\{1,\ldots,d-1\} \cdot d, j}\right) \bsS_{\{1,\ldots,d-1\} \cdot d, jj}^{-1/2}$. With this and equation \eqref{E:hFunction2a}, the partial derivative of the partial copula function with respect to a generic association parameter $\rho$ is
\begin{eqnarray}\label{E:AssocPartialCop1}
&&\frac{\partial}{\partial \rho}
C_d \left(u_1, \ldots, u_d \right) = \frac{\partial}{\partial \rho} \Phi_{d-1}\left(z_1 - \mu_{\{1, \ldots, d-1\} \cdot d, 1}, \ldots, z_{d-1} - \mu_{\{1, \ldots, d-1\} \cdot d, d-1} ; \bsS_{\{1, \ldots, d-1\} \cdot d} \right)  \notag \\
&& \ \ \ = \frac{\partial}{\partial \rho} \Phi_{d-1}\left(z_1^*, \ldots, z_{d-1}^* ; \mathbf{R}_{\{1, \ldots, d-1\} \cdot d} \right)\\
&& \ \ \ =
\sum_{k=1}^{d-1} h_{1,d-1}^{(k)}\left(z_1^*, \ldots, z_{d-1}^* ; \mathbf{R}_{\{1, \ldots, d-1\} \cdot d} \right)
\frac{\partial}{\partial \rho} z_k^* \notag\\
&& \ \ \ \ \ \ \ \ + \sum_{i<j}
h_{2,d-1}^{(ij)}\left(z_1^*, \ldots, z_{d-1}^* ; \mathbf{R}_{\{1, \ldots, d-1\} \cdot d} \right)
\frac{\partial}{\partial \rho}\mathbf{R}_{\{1, \ldots, d-1\} \cdot d,ij}\notag
\end{eqnarray}

\subsubsection*{Trivariate Case: Association Parameter Derivatives of $C_3$}

To illustrate, consider the trivariate case (more details available in \textit{Online Supplement} 2.1). Here, we seek to evaluate $\frac{\partial}{\partial \rho} \Phi_2 \left( z_1 - \mu_{12 \cdot 3, 1}, z_2 - \mu_{12 \cdot 3, 2} ; \bsS_{12 \cdot 3}\right).$

With the matrix $\bsS_{12 \cdot 3}$, to simplify notation, we define
$$\begin{array}{l}
\bsS_{12 \cdot 3,11} = 1-\rho_{13}^2 =\sigma_1^2 \\
\bsS_{12 \cdot 3,22} = 1-\rho_{23}^2 =\sigma_2^2 \\
\bsS_{12 \cdot 3,12} = \rho_{12} - \rho_{13}\rho_{23} =\rho_{X} \sigma_1\sigma_2.\\
\end{array}
$$
With this notation, we have
$$
\Phi_2 \left( z_1 - \mu_{12 \cdot 3, 1}, z_2 - \mu_{12 \cdot 3, 2} ; \bsS_{12 \cdot 3}\right)
=\Phi_2 \left( \frac{z_1 - \mu_{12 \cdot 3, 1}}{\sigma_1}, \frac{z_2 - \mu_{12 \cdot 3, 2} }{\sigma_2}; \rho_x\right)
=\Phi_2 \left( z_1^*, z_2^*; \rho_x\right)
$$
where $z_1^*=\left(z_1 - \mu_{12 \cdot 3, 1}\right)/\sigma_1$ and similarly for $z_2^*$.

From equation \eqref{E:hFunction1}, we have
\begin{eqnarray*}
h_{1,2}^{(1)}(x_1, x_2; \mathbf{R})
&=& \phi\left(x_1\right)\Phi\left(  \frac{x_2 - \mathbf{R}_{12} x_1  }{ \sqrt{1-\mathbf{R}_{12}^2}} \right) ,
\end{eqnarray*}

With this, we have
$$\begin{array}{ll}
&\frac{\partial}{\partial \rho} \Phi_2 \left( z_1 - \mu_{12 \cdot 3, 1}, z_2 - \mu_{12 \cdot 3, 2} ; \bsS_{12 \cdot 3}\right)
=
\frac{\partial}{\partial \rho} \Phi_2 \left( z_1^*, z_2^*; \rho_x\right)
\\
& \ \ \ \ \ =
h_{1,2}^{(1)}\left( z_1^*, z_2^*; \rho_x\right)\frac{\partial }{\partial \rho} z_1^*  +
h_{1,2}^{(1)}\left( z_2^*, z_1^*; \rho_x\right)\frac{\partial }{\partial \rho} z_2^*  +
\phi_2\left( z_1^*, z_2^*; \rho_x\right)\frac{\partial }{\partial \rho} \rho_x , \\
\end{array} .
$$
We also need
$$
\frac{\partial}{\partial \rho} z_1^* = \frac{\partial }{\partial \rho} \left(\frac{z_1 - \mu_{12 \cdot 3, 1}}{\sigma_1}\right)
= \frac{\partial }{\partial \rho} \left(\frac{z_1 - z_3 \rho_{13}}{\sqrt{1-\rho_{13}^2}}\right)
= \left\{\begin{array}{cl}
0 & \rho = \rho_{12} \\
\frac{z_1\rho_{13} - z_3 }{(1-\rho_{13}^2)^{3/2}} & \rho = \rho_{13} \\
0 & \rho = \rho_{23} \\
\end{array} \right.
$$

In the same way
$$
\frac{\partial}{\partial \rho} z_2^* = \frac{\partial }{\partial \rho} \left(\frac{z_2 - \mu_{12 \cdot 3, 2}}{\sigma_2}\right)
= \frac{\partial }{\partial \rho} \left(\frac{z_2 - z_3 \rho_{23}}{\sqrt{1-\rho_{23}^2}}\right)
= \left\{\begin{array}{cl}
0 & \rho = \rho_{12} \\
0 & \rho = \rho_{13} \\
\frac{z_2\rho_{23} - z_3 }{(1-\rho_{23}^2)^{3/2}} & \rho = \rho_{23}
\end{array} \right.
$$

Similarly,
$$
\frac{\partial}{\partial \rho} \rho_x = \frac{\partial }{\partial \rho} \left(\frac{\rho_{12} - \rho_{13}\rho_{23}}{\sigma_1 \sigma_2}\right)
= \frac{\partial }{\partial \rho} \left(\frac{\rho_{12} - \rho_{13}\rho_{23}}{(1-\rho_{13}^2)^{1/2} (1-\rho_{23}^2)^{1/2}}\right)
= \left\{\begin{array}{cl}
\frac{1}{(1-\rho_{13}^2)^{1/2} (1-\rho_{23}^2)^{1/2}} & \rho = \rho_{12} \\
\frac{\rho_{12}\rho_{13}-\rho_{23}}{(1-\rho_{13}^2)^{3/2} (1-\rho_{23}^2)^{1/2}} & \rho = \rho_{13} \\
\frac{\rho_{12}\rho_{23}-\rho_{13}}{(1-\rho_{13}^2)^{1/2} (1-\rho_{23}^2)^{3/2}}& \rho = \rho_{23}
\end{array} \right.
$$
This provides needed details to compute  $\frac{\partial}{\partial \rho} C_{3}$.

\subsubsection{Association Parameter Derivatives of $C_{d-1,d}$}

We next derive the partial derivative with respect association parameter of the copula function $C_{d-1,d} \left(u_1, \ldots, u_d \right)$. We start with
\begin{eqnarray*}
&&\frac{\partial}{\partial \rho}  C_{d-1,d} \left(u_1, \ldots, u_d \right) =
C \left(u_1, \ldots, u_{d-2} | u_{d-1}, u_d\right) \frac{\partial}{\partial \rho} c(u_{d-1}, u_d) \\
&& \ \ \ \ + \left(\frac{\partial}{\partial \rho} C \left(u_1, \ldots, u_{d-2} | u_{d-1}, u_d\right)\right) c(u_{d-1}, u_d) ,
\end{eqnarray*}
where
\begin{eqnarray*}
&& C \left(u_1, \ldots, u_{d-2} | u_{d-1}, u_d\right) \\
&& \ \ = \Phi_{d-2}\left(
z_1 - \mu_{\{1, \ldots, d-2\} \cdot \{d-1,d\},1}, \ldots, z_{d-2} - \mu_{\{1, \ldots, d-2\} \cdot \{d-1,d\},d-2} ; \bsS_{\{1, \ldots, d-2\} \cdot \{d-1,d\}}
\right).
\end{eqnarray*}
Further, as in equation \eqref{E:AssocPartialCop1}
\begin{eqnarray*}
&&\frac{\partial}{\partial \rho} \Phi_{d-2}\left(z_1 - \mu_{\{1, \ldots, d-2\} \cdot \{d-1,d\}}, \ldots, z_{d-2} - \mu_{\{1, \ldots, d-2\} \cdot \{d-1,d\}} ; \bsS_{\{1, \ldots, d-2\} \cdot \{d-1,d\}} \right) \\
&& \ \ \ = \frac{\partial}{\partial \rho} \Phi_{d-2}\left(z_1^*, \ldots, z_{d-2}^* ; \mathbf{R}_{\{1, \ldots, d-2\} \cdot \{d-1,d\}} \right) \notag \\
&& \ \ \ =
\sum_{k=1}^{d-2} h_{1,d-2}^{(k)}\left(z_1^*, \ldots, z_{d-2}^* ; \mathbf{R}_{\{1, \ldots, d-2\} \cdot \{d-1,d\}} \right) \frac{\partial}{\partial \rho} z_k^* \\
&& \ \ \ \ \ \ \ \ + \sum_{i<j}
h_{2,d-2}^{(ij)}\left(z_1^*, \ldots, z_{d-2}^* ; \mathbf{R}_{\{1, \ldots, d-2\} \cdot \{d-1,d\}} \right) \frac{\partial}{\partial \rho}\mathbf{R}_{\{1, \ldots, d-2\} \cdot \{d-1,d\},ij} ~.
\end{eqnarray*}
Here, the rescaled arguments are defined as $z_j^* = \left(z_j - \mu_{\{1, \ldots, d-2\} \cdot \{d-1,d\}, j}\right) \bsS_{\{1, \ldots, d-2\} \cdot \{d-1,d\}, jj}^{-1/2}$.

Derivatives of the bivariate density $c(u_{d-1}, u_d)$  are in Section \ref{S:BivariateGaussian}.

\subsubsection*{Trivariate Case: Association Parameter Derivatives of $C_{23}$}

To illustrate, for $d=3$, we have
$$
\frac{\partial}{\partial \rho}  C_{23} \left(u_1, u_2, u_3 \right) =
C \left(u_1 | u_{2}, u_3\right) \frac{\partial}{\partial \rho} c \left(u_{2}, u_3\right)
+ \left(\frac{\partial}{\partial \rho} C \left(u_1 | u_{2}, u_3\right)\right) c \left(u_{2}, u_3\right)
$$
where $C \left(u_1 | u_{2}, u_3\right) = \Phi\left( z_1 - \mu_{1 \cdot 23}; \sigma_{1 \cdot 23} \right).$ Further,
$$\begin{array}{ll}
\frac{\partial}{\partial \rho} C \left(u_1| u_2, u_3\right)&=
\frac{\partial}{\partial \rho} \Phi\left(z_1 - \mu_{1 \cdot 23}; \sigma_{1 \cdot 23}\right) \nonumber \\
&=
\phi\left( \frac{z_1 - \mu_{1 \cdot 23}} {\sqrt{\sigma_{1 \cdot 23}}}\right) \frac{\partial}{\partial \rho}
\left( \frac{z_1 - \mu_{1 \cdot 23}} {\sqrt{\sigma_{1 \cdot 23}}}\right) \nonumber \\
&=
- \phi\left( \frac{z_1 - \mu_{1 \cdot 23}} {\sqrt{\sigma_{1 \cdot 23}}}\right)
\frac{1}{\sigma_{1 \cdot 23}^{3/2}}
\left(
 \sigma_{1 \cdot 23} \frac{\partial}{\partial \rho} \mu_{1 \cdot 23} +
\frac{1}{2}(z_1 - \mu_{1 \cdot 23}) \frac{\partial}{\partial \rho} \sigma_{1 \cdot 23}
\right) .
\end{array} $$

Recall
$\frac{\partial}{\partial \rho} \bsS^{-1} = -\bsS^{-1} \left(\frac{\partial}{\partial \rho} \bsS\right) \bsS^{-1}$. Now, with $\mu_{1 \cdot 23} = \left(\begin{array}{cc} \rho_{12} & \rho_{13}\end{array}\right) \left(\begin{array}{cc} 1& \rho_{23} \\ \rho_{23}&1\end{array}\right)^{-1} \left(\begin{array}{c} z_2  \\ z_3 \end{array}\right)$, we have

$$
\frac{\partial}{\partial \rho} \mu_{1 \cdot 23}
= \left\{\begin{array}{cl}
\left(\begin{array}{cc} 1 & 0 \\ \end{array}\right)
\left(\begin{array}{cc} 1& \rho_{23} \\ \rho_{23}&1  \\ \end{array}\right)^{-1}
\left(\begin{array}{c} z_2  \\ z_3   \\ \end{array}\right)& \rho = \rho_{12} \\
\left(\begin{array}{cc} 0 & 1 \\ \end{array}\right)
\left(\begin{array}{cc} 1& \rho_{23} \\ \rho_{23}&1  \\ \end{array}\right)^{-1}
\left(\begin{array}{c} z_2  \\ z_3   \\ \end{array}\right) & \rho = \rho_{13} \\
\left(\begin{array}{cc} \rho_{12} & \rho_{13} \\ \end{array}\right)
\left(\begin{array}{cc} 1& \rho_{23} \\ \rho_{23}&1  \\ \end{array}\right)^{-1}
\left(\begin{array}{cc} 0& 1 \\ 1&0  \\ \end{array}\right)
\left(\begin{array}{cc} 1& \rho_{23} \\ \rho_{23}&1  \\ \end{array}\right)^{-1}
\left(\begin{array}{c} z_2  \\ z_3   \\ \end{array}\right)& \rho = \rho_{23}
\end{array} \right.
$$

Further, with $\sigma_{1 \cdot 23} = 1-\left(\begin{array}{cc} \rho_{12} & \rho_{13} \\  \end{array}\right)\left(\begin{array}{cc} 1& \rho_{23} \\ \rho_{23}&1  \\ \end{array}\right)^{-1}\left(\begin{array}{c} \rho_{12} \\ \rho_{13} \\  \end{array}\right)$, we have

$$
\frac{\partial}{\partial \rho} \sigma_{1 \cdot 23}
= \left\{\begin{array}{cl}
- 2
\left(\begin{array}{cc} 1 &  0 \\ \end{array}\right)
\left(\begin{array}{cc} 1& \rho_{23} \\ \rho_{23}&1  \\ \end{array}\right)^{-1}
\left(\begin{array}{c} \rho_{12} \\ \rho_{13} \\ \end{array}\right)& \rho = \rho_{12} \\
- 2
\left(\begin{array}{cc} 0 &  1 \\ \end{array}\right)
\left(\begin{array}{cc} 1& \rho_{23} \\ \rho_{23}&1  \\ \end{array}\right)^{-1}
\left(\begin{array}{c} \rho_{12} \\ \rho_{13} \\ \end{array}\right) & \rho = \rho_{13} \\
-
\left(\begin{array}{cc} \rho_{12} &  \rho_{13} \\ \end{array}\right)
\left(\begin{array}{cc} 1& \rho_{23} \\ \rho_{23}&1  \\ \end{array}\right)^{-1}
\left(\begin{array}{cc} 0& 1 \\ 1&0  \\ \end{array}\right)
\left(\begin{array}{cc} 1& \rho_{23} \\ \rho_{23}&1  \\ \end{array}\right)^{-1}
\left(\begin{array}{c} \rho_{12} \\ \rho_{13} \\ \end{array}\right)& \rho = \rho_{23}
\end{array} \right.
$$
This provides needed details to compute  $\frac{\partial}{\partial \rho} C_{23}$.

\end{document}